\documentclass{aa}
\usepackage{graphics,epsfig,txfonts,xspace}
\usepackage{natbib}
\usepackage[usenames]{color}
\bibpunct{(}{)}{;}{a}{}{,}

\newcommand{\ergcm}[1]{$\times 10^{#1}$ erg cm$^{-2}$ s$^{-1}$}

\newcommand{\ergs}[1]{$\times 10^{#1}$ erg s$^{-1}$}

\newcommand{\hcm}[1]{$\times 10^{#1}$ cm$^{-2}$}

\newcommand{\expo}[1]{$\times 10^{#1}$}
\newcommand{\oexpo}[1]{$10^{#1}$}

\newcommand{\nh}{N$_{\rm H}$}

\newcommand{\ct}{cts s$^{-1}$}

\newcommand{\Halp}{H${\alpha}$\xspace}
\newcommand{\ltsima}{$\buildrel < \over \sim$}
\newcommand{\lsim}{\lower.5ex\hbox{\ltsima}}
\newcommand{\gtsima}{$\buildrel > \over \sim$}
\newcommand{\gsim}{\lower.5ex\hbox{\gtsima}}
\newcommand{\xmm}{XMM-Newton\xspace}

\newcommand{\xmmp}{\hbox{XMMU\,J004814.0-732204}\xspace}

\begin{document}
 
\title{The XMM-Newton survey of the Small Magellanic Cloud:\\ 
       Discovery of the 11.866 s Be/X-ray binary pulsar \xmmp (SXP11.87)}
\author{R.~Sturm\inst{1} \and F.~Haberl\inst{1} \and M.J. Coe\inst{2} 
        \and E.S. Bartlett\inst{2} 
	\and D.A.H. Buckley\inst{3}
	\and R.H.D. Corbet\inst{4}
	\and M. Ehle\inst{5} 
	\and M.D. Filipovi{\'c}\inst{6}
	\and D. Hatzidimitriou\inst{7,8} 
	\and S. Mereghetti\inst{9} 
	\and N. La Palombara\inst{9} 
	\and W. Pietsch\inst{1} 
	\and A. Tiengo\inst{9} 
	\and L.J. Townsend\inst{2}
	\and A. Udalski\inst{10}
	}

\titlerunning{A 11.87 s Be/X-ray binary pulsar in the SMC}
\authorrunning{Sturm et al.}

\institute{Max-Planck-Institut f\"ur extraterrestrische Physik,
           Giessenbachstra{\ss}e, 85748 Garching, Germany
	   \and
           School of Physics and Astronomy, University of Southampton, Highfield, Southampton SO17 1BJ, United Kingdom
	   \and
	   South African Astronomical Observatory, PO Box 9, Observatory 7935, Cape Town, South Africa
	   \and
	   University of Maryland, Baltimore County, Mail Code 662, NASA Goddard Space Flight Center, Greenbelt, MD 20771, USA
           \and
           XMM-Newton Science Operations Centre, ESAC, ESA, PO Box 78, 28691 Villanueva de la Ca\~{n}ada, Madrid, Spain
	   \and
           University of Western Sydney, Locked Bag 1797, Penrith South DC, NSW1797, Australia
	   \and
	   Department of Astrophysics, Astronomy and Mechanics, Faculty of Physics, University of Athens, Panepistimiopolis, GR15784 Zografos, Athens, Greece
	   \and
	   Foundation for Research and Technology Hellas, IESL, Greece
	   \and
	   INAF, Istituto di Astrofisica Spaziale e Fisica Cosmica Milano, via E. Bassini 15, 20133 Milano, Italy
           \and
           Warsaw University Observatory, Aleje Ujazdowskie 4, 00-478 Warsaw, Poland
	   }

\date{Received 21 September 2010 / Accepted 6 November 2010}
 
\abstract{} 
         {One of the goals of the XMM-Newton survey of the Small Magellanic Cloud is the study of the Be/X-ray binary population.
          During one of our first survey observations a bright new transient $-$ \xmmp $-$ was discovered.
         }
	 {We present the analysis of the EPIC X-ray data together with optical observations, 
          to investigate the spectral and temporal characteristics of \xmmp.}
         {We found coherent X-ray pulsations in the EPIC data with a period of $(11.86642\pm 0.00017)$ s. The X-ray spectrum can be modelled by an 
	  absorbed power-law with indication for a soft excess. Depending on the modelling of the soft X-ray spectrum, the photon index ranges 
	  between 0.53 and 0.66. We identify the optical counterpart as a B = 14.9mag star 
	  which was monitored during the MACHO and OGLE-III projects. The optical light curves show regular outbursts
	  by $\sim$0.5 mag in B and R and up to 0.9 mag in I which repeat with a time scale of about 1000 days. 
	  The OGLE-III optical colours of the star are consistent with an early B spectral type.
	  An optical spectrum obtained at the 1.9\,m telescope of the South African Astronomical Observatory in December 2009 shows \Halp emission 
	  with an equivalent width of 3.5 $\pm$ 0.6\,\AA.}
         {The X-ray spectrum and the detection of pulsations suggest that \xmmp is a new high mass X-ray binary pulsar in the SMC. 
	  The long term variability and the \Halp emission line in the spectrum of the optical counterpart identify it as a Be/X-ray binary system.}

\keywords{galaxies: individual: Small Magellanic Cloud --
          galaxies: stellar content --
          stars: emission-line, Be -- 
          stars: neutron --
          X-rays: binaries}
 
\maketitle
 
\section{Introduction}

The Small Magellanic Cloud (SMC) hosts an extraordinary high number of about 80 known Be/X-ray binary systems, compared to the $\sim$70 
known in the Galaxy \citep[as of 2006,][]{2006A&A...455.1165L} which is a factor of $\sim$100 more massive than the SMC. 
Be/X-ray binaries are a subclass of high mass X-ray binaries containing an early type Be donor star 
with equatorial mass ejection, and an accreting neutron star (NS).
Due to the non-spherical and time variable mass ejection these systems show up as X-ray transients, when the NS crosses the disk during 
the periastron passage, leading to enhanced matter accretion for a few days (type I outbursts). 
Longer outbursts lasting several weeks (type II) are thought to be caused by expansion of the circumstellar disk \citep[see e.g. ][]{2001A&A...377..161O}.

One of the aims of the \xmm \citep{2001A&A...365L...1J} large program SMC survey \citep{2008xng..conf...32H} is the ongoing study of 
the Be/X-ray binary population of the SMC, which can be used as a star formation tracer for $\sim50$ (30-70) Myr old populations \citep{2010ApJ...716L.140A}.
In this paper we present the analysis of X-ray and optical data from the newly discovered X-ray pulsar \xmmp.

\section{Observations and data reduction}

The new transient was discovered on 2009 Oct. 03, during observation 13 (observation ID 0601211301) of the \xmm large program SMC survey. 
The source was located near the border of CCD\,1 (partly spread onto CCD\,4) of the EPIC-pn instrument \citep{2001A&A...365L..18S} 
and on CCD\,2 of EPIC-MOS2 \citep{2001A&A...365L..27T}. There are no MOS1 data for this source because it was located on CCD\,6, 
which is switched off since \xmm\ revolution 961.
The soft proton background was at a very low level during the whole observation. Therefore, no background screening was necessary,
resulting in net exposure times of 30779 s and 32368 s for EPIC-pn and EPIC-MOS2, respectively.

We used XMM-Newton SAS 10.0.0 \footnote{Science Analysis Software (SAS), http://xmm.esac.esa.int/sas/} to process the data. 
We identified sources in the field of view (FoV) for astrometric bore-sight correction by comparison with the Magellanic Clouds Photometric 
Survey of \citet{2002AJ....123..855Z}, obtaining a shift of $\Delta$RA=-0.15\arcsec\ and $\Delta$Dec=-1.23\arcsec.
The corrected position of the transient as found by {\tt emldetect} is 
R.A. = 00$^{\rm h}$48$^{\rm m}$14\fs07 and Dec. = --73\degr22\arcmin04\farcs4 (J2000.0), 
with a statistical error of 0.06\arcsec\ and a systematic uncertainty of $\sim$1\arcsec\ (1 $\sigma$ confidence for both cases).

\begin{figure}
  \resizebox{\hsize}{!}{\includegraphics[angle=0,clip=]{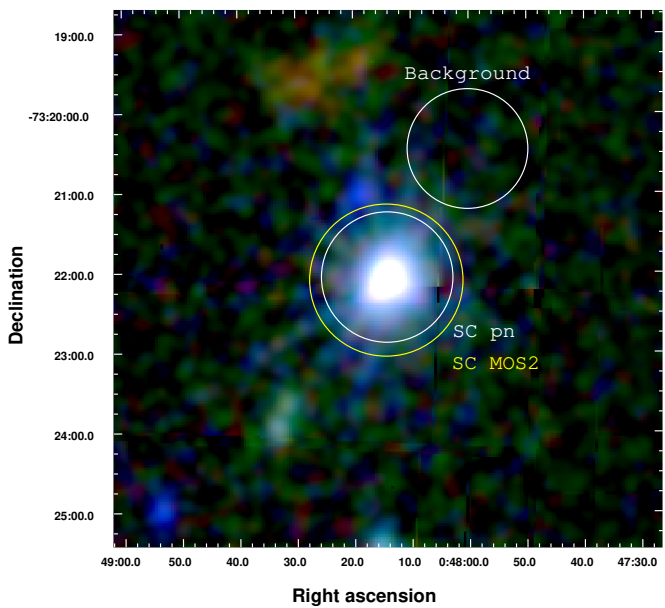}}
  \caption{EPIC colour image of \xmmp combining pn and MOS data. The red, green and blue colours represent the X-ray intensities in the 
           0.2$-$1.0, 1.0$-$2.0 and 2.0$-$4.5 keV energy bands. Circles indicate the extraction regions 
	   (with radii of 49\arcsec\ and 57\arcsec\ for pn and MOS2 source regions and 45\arcsec\ for the background). }
  \label{fig:ima}
\end{figure}

For the extraction of EPIC spectra, we selected single-pixel events from the EPIC-pn data ({\tt PATTERN = 0}) and single to quadruple 
events with {\tt PATTERN$\le$12} from EPIC-MOS2 data, both with {\tt FLAG = 0}.
The SAS task {\tt eregionanalyse} was used to determine circular source extraction regions by optimizing the signal to noise ratio
as shown in Fig.~\ref{fig:ima}. We ensured that the source extraction region has a distance of $>$10\arcsec\ to other detected sources.
For the background extraction region, we chose a circle in an area free of point sources and on the same CCD as the source for both instruments.
The EPIC-pn and EPIC-MOS2 spectra contain 9286 and 8054 background subtracted counts, respectively, and were binned to a minimum signal-to-noise ratio of 5 for each bin. 
For the timing analysis, we used also double-pixel events for EPIC-pn. 
To increase the statistics for the timing analysis we also generated a merged event list from both instruments, containing 25945 cts (source + background).

\section{X-ray data analysis and results}

\subsection{Spectral analysis of the X-ray data}

\begin{figure}
  \resizebox{\hsize}{!}{\includegraphics[angle=-90,clip=]{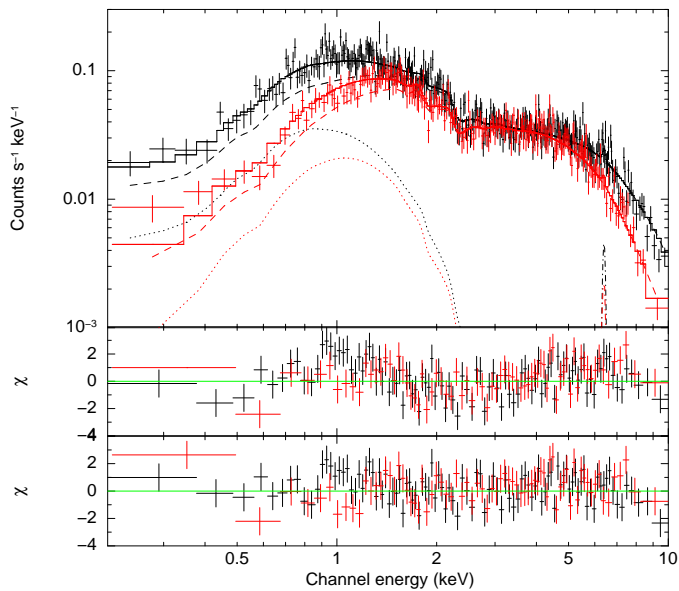}}
  \caption{EPIC spectra of \xmmp. The top panel shows the EPIC-pn (black) and EPIC-MOS2 (red) spectra together with the best-fit model (solid line) 
           of an absorbed power-law (dashed line) plus black-body (dotted line) and iron fluorescent line (dash-dotted line). 
	   The residuals (for better comparison they are re-binned by an additional factor of three) are plotted for 
	   this model (bottom panel) and for the best-fit single power-law model (middle panel).}
  \label{fig:spec}
\end{figure}

We used {\tt XSPEC} \citep{1996ASPC..101...17A} version 12.5.0x for spectral fitting.
The two EPIC spectra were fitted simultaneously with a common set of spectral model parameters, only a relative normalisation factor was allowed 
to vary to account for instrumental differences.
The spectrum (Fig.~\ref{fig:spec}) was modelled first with an absorbed power-law. We fixed the Galactic photo-electric absorption
at a column density  of N$_{\rm H{\rm , GAL}}$ = 6\hcm{20} with abundances according to \citet{2000ApJ...542..914W},
whereas the SMC column density was a free parameter with abundances for elements heavier than Helium fixed at 0.2.
The best-fit parameters are summarised in Table~\ref{tab-spectra} where errors denote 90\% confidence ranges. 

The extraction for the EPIC-pn spectrum is hampered by the CCD gap cutting the extraction region. The missing area is 
taken into account in the calculation of the effective area by {\tt arfgen}. However, we noticed that when using the default 
spatial resolution (parameter badpixelresolution=2.0\arcsec) the flux derived from the EPIC-pn spectrum is higher by ($21\pm3$)\% 
than compared to MOS. Using badpixelresolution=1.0\arcsec\ reduces the flux discrepancy to 7\%, which is within the expected 
systematic uncertainties in the presence of gaps. 
Extracting the EPIC-pn spectrum from a smaller source region with radius 6\arcsec, so that the complete source region is placed on CCD\,1,
yields a flux that only differs by $\sim$1\% from the MOS2 value. The spectral shape is not affected by the CCD gap, but the number of 
source counts for the smaller extraction region is a factor of two lower.

In principle, this fit is formally acceptable and additional components are not required.
However, soft excesses and fluorescent emission from iron are known to contribute to the X-ray emission of some Be X-ray binaries 
\citep[e.g.][]{2008A&A...491..841E,2009A&A...505..947L,2004ApJ...614..881H}.
To investigate these possibilities we first added a black-body emission component to the model (Table~\ref{tab-spectra}).
This component contributes $\sim$2\% to the observed flux and $\sim$3\% to the absorption corrected 
luminosity. For the bolometric luminosity we obtained $(1.40\pm 0.35) \times 10^{35}$ erg s$^{-1}$.
Compared to the single power-law the reduced $\chi^2$ improved from 1.09 to 1.01, which corresponds to a F-test chance 
probability of $2.6 \times 10^{-10}$ and formally proves the significance of this component 
\citep[but see][ for limitations of the F-test]{2002ApJ...571..545P}.
An additional emission line with fixed energy at 6.4 keV and unresolved line width (fixed at 0), 
yielded a line flux of $4.6\pm 4.0\times 10^{-6}$ photons cm$^{-2}$ s$^{-1}$ 
corresponding to an equivalent width given in Table~\ref{tab-spectra}.
Substituting the 6.4 keV line by a 6.7 keV line for ionized \ion{Fe}{XXV} resulted in an upper limit for the equivalent width of 41 eV.

If we replace the black-body component by a multi-temperature disk black-body model (diskbb in {\tt XSPEC}), we derive
a lower limit for the inner disk radius of $R_{\rm in}=5.9^{+3.5}_{-2.2}$ km (for a disk inclination of $\Theta=0$ with $R_{\rm in} \propto 1/\sqrt{cos\Theta}$).
Following \citet{2004ApJ...614..881H} to estimate the inner disk radius we infer $R_{\rm in} = \sqrt{L_{\rm X}/(4 \pi \sigma T^4)}= 39$ km.

\begin{table*}
\caption[]{Spectral fit results.}
\begin{center}
\begin{tabular}{lcccccccc}
\hline\hline\noalign{\smallskip}
\multicolumn{1}{l}{Model$^{(1)}$} &
\multicolumn{1}{c}{SMC \nh} &
\multicolumn{1}{c}{$\gamma$} &
\multicolumn{1}{c}{kT} &
\multicolumn{1}{c}{R$^{(2)}$} &
\multicolumn{1}{c}{EW$_{\rm Fe}$} &
\multicolumn{1}{c}{Flux$^{(3)}$} &
\multicolumn{1}{c}{L$_{\rm x}^{(4)}$} &
\multicolumn{1}{c}{$\chi^2/{\rm dof}$} \\
\multicolumn{1}{c}{} &
\multicolumn{1}{c}{[\oexpo{21}cm$^{-2}$]} &
\multicolumn{1}{c}{} &
\multicolumn{1}{c}{[eV]} &
\multicolumn{1}{c}{[km]} &
\multicolumn{1}{c}{[eV]} &
\multicolumn{1}{c}{[erg cm$^{-2}$ s$^{-1}$]} &
\multicolumn{1}{c}{[erg s$^{-1}$]} &
\multicolumn{1}{c}{} \\

\noalign{\smallskip}\hline\noalign{\smallskip}
 PL               & 1.72$\pm$0.25  & 0.66$\pm$0.03 & --         & --               & --           &(9.0$\pm$0.3)\expo{-12} & 4.0\expo{36} & 570/523 \\
 PL+BB            & 2.32$\pm$0.44  & 0.52$\pm$0.05 & 279$\pm$43 & 12.2$\pm$1.4     & --           &(9.3$\pm$0.5)\expo{-12} & 4.2\expo{36} & 524/521 \\
 PL+BB+Fe-line    & 2.34$\pm$0.45  & 0.53$\pm$0.05 & 277$\pm$40 & 12.2$\pm$1.4     & 35$\pm$30    &(9.6$\pm$0.5)\expo{-12} & 4.2\expo{36} & 520/520 \\
 PL+DiskBB        & 2.99$\pm$0.54  & 0.52$\pm$0.06 & 383$\pm$97 & $>$5.9$\pm$3.5   & --           &(9.3$\pm$0.7)\expo{-12} & 4.2\expo{36} & 528/521 \\

\noalign{\smallskip}\hline
\end{tabular}
\end{center}
$^{(1)}$ For definition of spectral models see text. 
$^{(2)}$ Radius of the emitting area (for BB) or inner disk radius (DiskBB, for the definition see text).
$^{(3)}$ Observed 0.2-10.0 keV flux.
$^{(4)}$ Source intrinsic X-ray luminosity in the 0.2-10.0 keV band (corrected for absorption)
for a distance to the SMC of 60 kpc \citep{2005MNRAS.357..304H}.
\label{tab-spectra}
\end{table*}

\subsection{Timing analysis of the X-ray data}

\begin{figure}
  \resizebox{\hsize}{!}{\includegraphics[angle=-90,clip=]{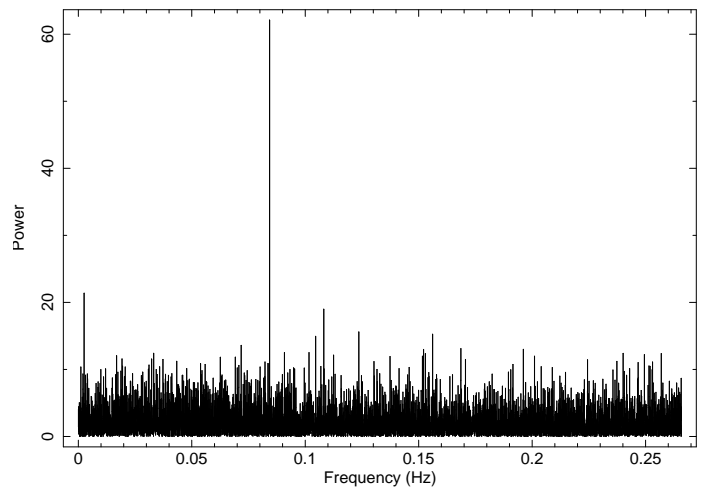}}
  \caption{Power density spectrum created from the merged EPIC-pn and EPIC-MOS2 data in the 0.2-10.0 keV energy band. 
           The time binning of the input light curve is 1.882 s. }
  \label{fig:psd}
\end{figure}

\begin{figure}
  \resizebox{\hsize}{!}{\includegraphics[angle=0,clip=]{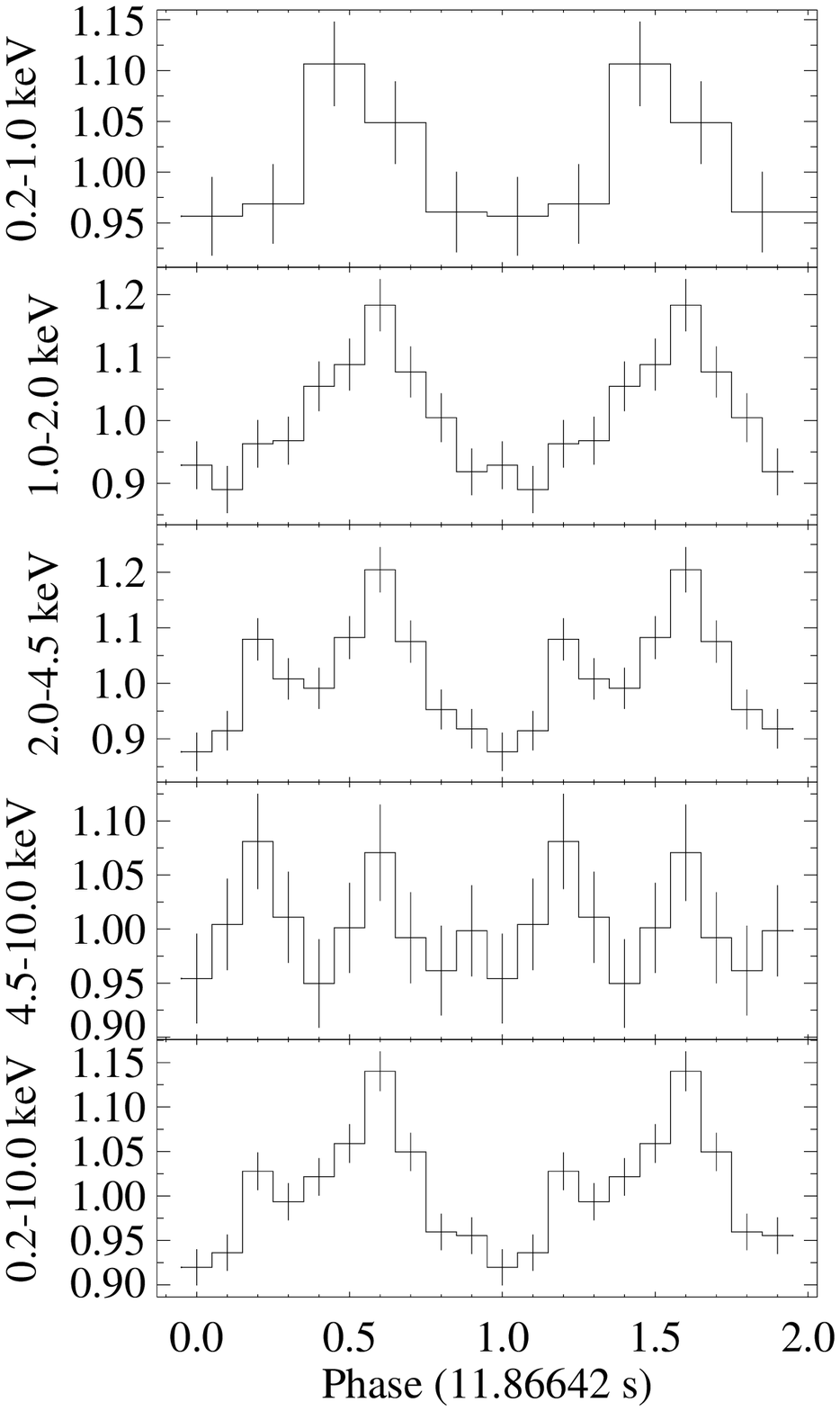}\includegraphics[angle=0,clip=]{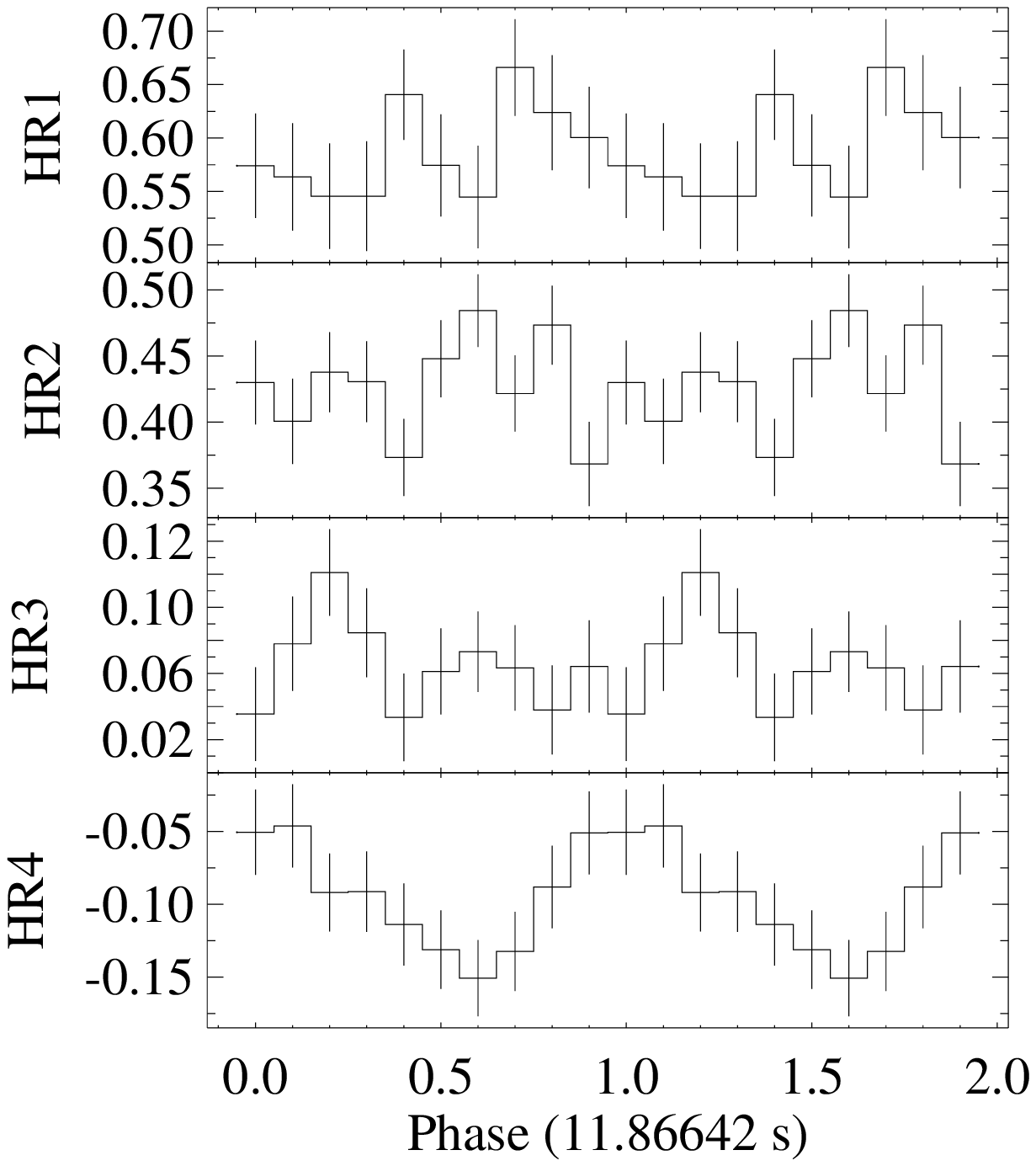}}
  \caption{Left: Pulse profiles obtained from the merged EPIC data in different energy bands 
           (for better statistics the first two standard energy bands were combined in the top panel, the bottom panel shows all five energy bands combined). 
	   The profiles are background-subtracted and normalized to the 
           average count rate (0.116, 0.228, 0.249, 0.207 and 0.801 \ct, from top to bottom. Right: Hardness ratios as a function of pulse phase derived 
	   from the pulse profiles in two neighbouring standard energy bands}
  \label{fig:pp}
\end{figure}

We corrected the event arrival times to the solar system barycentre using the SAS task {\tt barycen} and searched for periodicities 
in the X-ray light curves using fast Fourier transform (FFT) and light curve folding techniques.
The power density spectra derived from light curves in various energy bands from both EPIC instruments showed a periodic signal at 0.084~Hz. 
To increase the signal to noise ratio, we then created light curves from the merged event list of EPIC-pn and EPIC-MOS2 (delimited to common time intervals).
Figure~\ref{fig:psd} shows the inferred power density spectrum from the 0.2-10.0 keV energy band with the clear peak at a frequency of 0.084~Hz.
Following \citet{2008A&A...489..327H} we used a Bayesian periodic signal detection method \citep{1996ApJ...473.1059G} to determine the pulse period 
with 1$\sigma$ error to $(11.86642\pm 0.00017)$ s. 
The pulse profiles folded with this period in the EPIC standard energy bands (0.2-0.5 keV, 0.5-1.0 keV, 1.0-2.0 keV, 2.0-4.5 keV and 4.5-10 keV) 
are plotted in Fig.~\ref{fig:pp} together with hardness ratios derived from the pulse 
profiles in two adjacent energy bands (HR$_{i}$ = (R$_{i+1}$ $-$ R$_{i}$)/(R$_{i+1}$ + R$_{i}$) with R$_{i}$ denoting the background-subtracted 
count rate in energy band i (with i from 1 to 4). Assuming a sinusoidal pulse profile, we determined a pulsed fraction of $(7.5 \pm 1.0)$\% for the 0.2$-$10.0 keV band.
The profiles suggest some evolution from a single-peaked to a double-peaked structure with increasing energy, causing the variations in 
hardness ratios HR3 and HR4. A strong dependence of the pulse profiles on energy \citep[e.g. ][]{2003ApJ...584..996W,2008A&A...489..327H} 
and luminosity \citep[e.g. ][]{1997ApJS..113..367B} is seen from a number of high mass X-ray binaries.

\subsection{Long-term X-ray variability}

The position of \xmmp was covered in two previous \xmm observations on 2000 Oct. 15 (ObsID: 0110000101) and 2007 Apr. 11 (ObsID: 0404680301) 
with a background-screened net exposure of 21.6 ks and 17.6 ks, respectively. In the latter observation the source position was only covered by the MOS2 FoV.
In both observations, no source was detected above a likelihood threshold of 6. Using sensitivity maps we derived 3$\sigma$ upper limits of 
2.5$\times 10^{-3}$ \ct\ and 2.7$\times 10^{-3}$ \ct, respectively.
Assuming the same spectrum as during the outburst, this corresponds to a flux limit of 1.7\ergcm{-14} (from Oct. 2000, measured by EPIC-pn) and 
6.1\ergcm{-14} (Apr. 2007, EPIC-MOS2) in the $0.2-10.0$ keV band and to luminosity limits of 7.6\ergs{33} and 2.7\ergs{34}, respectively.

Also in a Chandra \citep{2000SPIE.4012....2W} ACIS-I observation (Observation ID 2945) on 2002 Oct. 2 this position was covered with a 11.8 ks exposure, and no source was detected.
We used the CIAO (Version 4.2) task {\tt aprates} to estimate a 3$\sigma$ upper limit of 5.1$\times 10^{-4}$ cts s$^{-1}$.
Assuming the same spectrum as above, this corresponds to a flux limit of 1.6\ergcm{-14} in the $0.2-10.0$ keV band (luminosity of 7.1\ergs{33}).

The upper limits from the \xmm and Chandra observations show that \xmmp increased in brightness at least by a factor of 560 during its outburst. 

RXTE monitoring of the SMC has been carried out for nearly a decade \citep{2008ApJS..177..189G} and \xmmp has frequently fallen within the 
pointing direction of the telescope, often at a collimator response of $\ge$60\%. Unfortunately, the time of the \xmm detection on MJD 55107 
falls into the gap in the RXTE monitoring (MJD 55080 - 55138) when the spacecraft was temporarily disabled, so there is no simultaneous 
RXTE coverage. Within the envelope of the last 9.5 years there are no possible detections of this source in the periods of time 
(approximately 6.5 years) when the source was above a 0.6 collimator response. The RXTE pulse monitoring is sensitive to detections 
with a pulsed amplitude in excess of 0.15-0.2 cts/s/PCU. This approximately translates into $\sim$0.5-1.0 cts/s/PCU - depending on 
the collimator response and pulsed fraction (which is not very high for \xmmp) - or a luminosity limit of $\sim$2\ergs{36} for a source in the SMC.

\section{Optical data}

\subsection{Identification of the optical counterpart}

\begin{figure}
  \resizebox{\hsize}{!}{\includegraphics[angle=0,clip=]{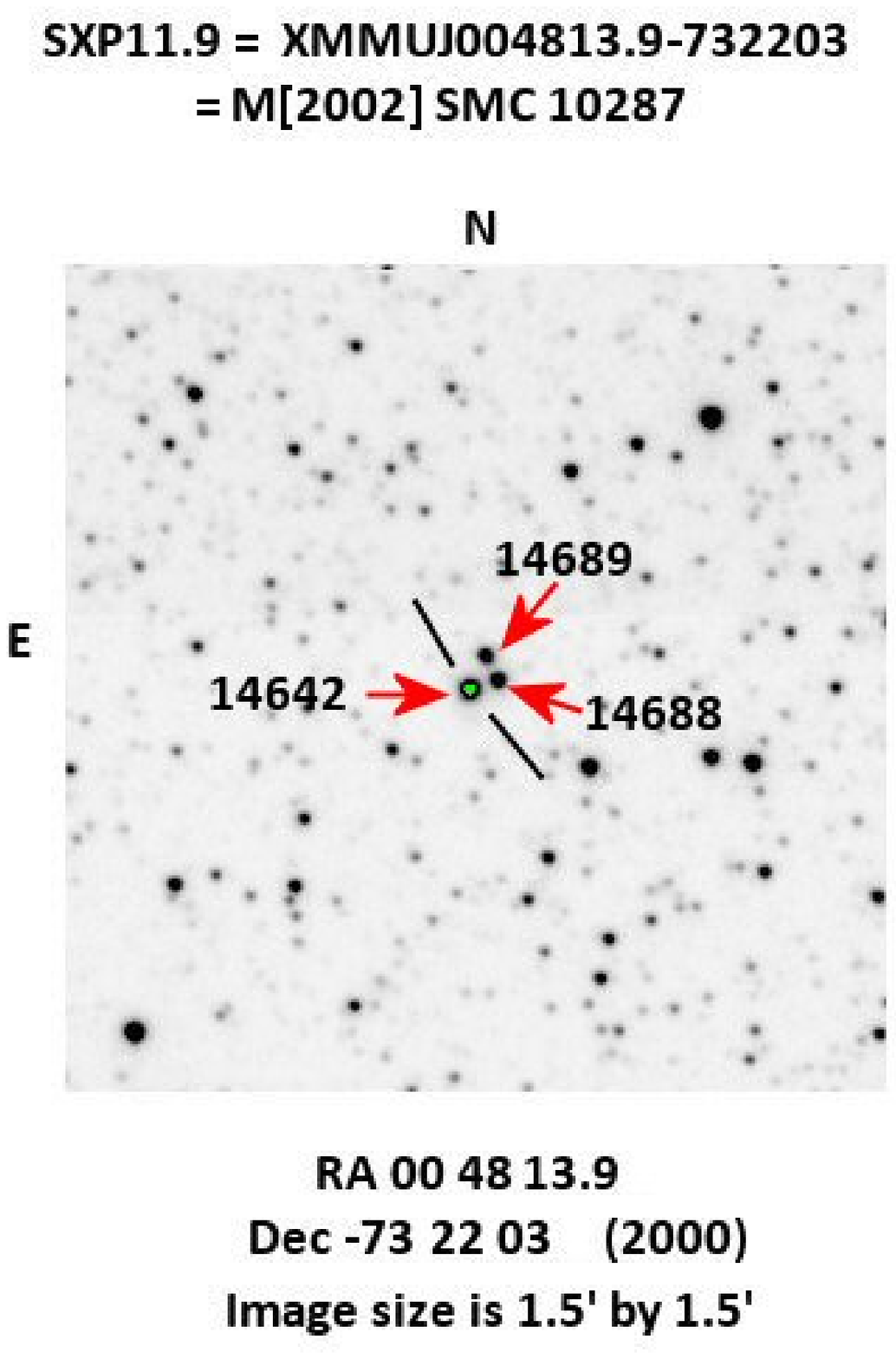}}
  \caption{Finding chart of SXP11.87. The I band image from OGLE-II shows the 3 close objects near the X-ray position marked with their OGLE-III 
           identification (arrows). The two lines further mark the likely counter part.
           The image size is 1\farcm 5 by 1\farcm 5. }
  \label{fig:fc}
\end{figure}

Searching optical catalogues of \citet{2002AJ....123..855Z}, MACHO and OGLE, we found three stars which are located within the 3$\sigma$ error radius 
around the \xmm position. Their positions and magnitudes from \citet{2002AJ....123..855Z} and their OGLE-II and MACHO entries are listed in 
Table~\ref{tab:opt-cp}. A finding chart produced from OGLE-III data is shown in Fig.~\ref{fig:fc}. 

The star closest to the X-ray position (OGLE-III 14642) has colours and brightness consistent with an early B star.
Its position on the U-B vs. B-V diagram of Be stars \citep[e.g. Fig.~1 of ][]{1979AJ.....84.1713F} is also entirely consistent with it being a Be star. 
The same holds for the reddening-free Q-index of -0.85 \citep{1955ApJ...122..142J,2007AJ....134.2474M}.
This candidate also appears as number 10287 in the survey list of \citet{2002ApJS..141...81M}. 
The (B-V) colour index from that catalogue is (B-V)=-0.12$\pm$0.01. 
Correcting for an extinction to the SMC of E(B-V)=0.09 \citep[][ also used for our spectral type estimates hereafter]{1991A&A...246..231S} 
gives an intrinsic colour of (B-V)=-0.21$\pm$0.01. 
From \citet{1994MNRAS.270..229W} this indicates a spectral type in the range B1.5V - B2.5V - typical of optical counterparts to 
Be/X-ray binaries in the SMC \citep{2008MNRAS.388.1198M}. However, care must always be taken when interpreting colour information as a 
spectral type in systems that clearly have circumstellar disks contributing some signal to the B and V bands.

Optical photometry was performed at the Faulkes Telescope South (FTS) on 
25 November 2009 (MJD 55160). The telescope is located at Siding Spring, Australia and is a 2m, fully autonomous, robotic 
Ritchey-Chr\`etien reflector 
on an alt-azimuth mount. The telescope employs a Robotic Control System (RCS). The telescope 
was used in Real Time Interface mode for the observation of \xmmp. All the observations were pipeline-processed (flat-fielding
and de-biasing of the images). The I-band magnitude of the optical counterpart was determined to be 15.30$\pm$0.02 mag by 
comparison with several other nearby stars on the same image frame and in the OGLE database. 
These comparison stars have not exhibited any significant variability in the last 8 years of OGLE monitoring. 

\begin{figure}
  \resizebox{\hsize}{!}{\includegraphics[angle=-90,clip=]{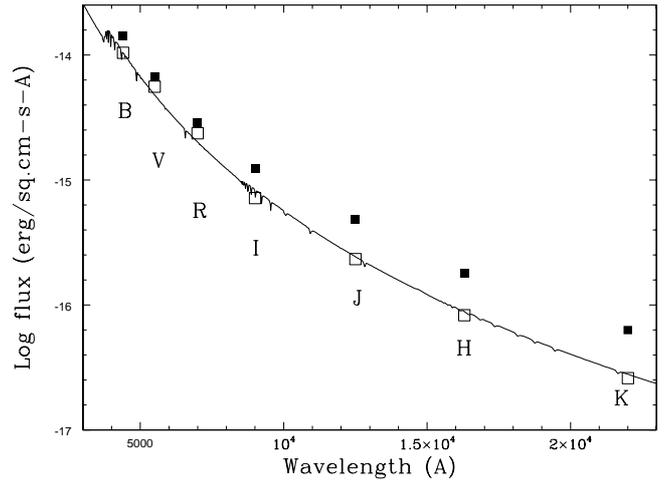}}
  \caption{Combined optical-IR flux for our counterpart OGLE-III 14642 at two epochs. (a) a historical data set (1999-2002) - solid symbols; 
           (b) data set from the time of outburst (Nov-Dec 2009) - open symbols. See text for details of the observations. 
	   Both data sets are compared to a Kurucz model atmosphere for a B2V star in which this stellar model has been normalised 
	   to the outburst B band point.}
  \label{fig:opt-sed}
\end{figure}

Figure~\ref{fig:opt-sed} shows optical and IR photometry of OGLE-III 14642 taken at the different epochs - 
see Table~\ref{tab:opt-mag} for the actual values.
The earliest optical data come from \citet{2002ApJS..141...81M} and were recorded on 1999 Jan. 8. 
These data are combined with IR measurements taken on 2002 Aug. 31 with the Sirius camera on the 1.5m IRSF telescope 
in South Africa \citep{2007PASJ...59..615K}. 
Also included is the OGLE I band measurement taken simultaneously with the Sirius IR data set. 
These early data are compared to a B, V, R \& I photometric data set recorded on 2009 Nov. 25 from the FTS. 

For comparison, a stellar atmosphere model \citep{1979ApJS...40....1K} representing a B2V star 
(T$_{\rm eff}$ = 22,000 K and log(g) = 4.0) is also shown where the model has been normalised 
to the most recent B band measurement. It is very clear that the recent
data taken around the time of the \xmm detection represent
the source in a much lower activity state than the earlier
data. Furthermore, the shape of the model atmosphere indicates
clear evidence for the presence of a significant IR excess in the past, 
almost certainly arising from the circumstellar disk of the system. 
But in late 2009 the disk had diminished significantly, to the extent that there 
is little in the way of IR excess - this is supported by the very weak H$\alpha$ 
emission (see Section 4.3 below).

The star OGLE-III 14688 is probably a red (K to M) giant from its OGLE colours. The light curve shows small variations of the order of 0.1 mag in 
the I-band, but no evidence for any coherent fluctuations. 
There is an object from the Two Micron All Sky Survey \citep{2006AJ....131.1163S} $-$ 2MASS\,J00481347-7322030 $-$ which is closest to the position of 
this star, with J=14.76, H=14.13, K=14.04, J-H=0.63 and H-K=0.09, which rather points towards a K2 star.

The third object (OGLE-III 14689) is likely a late B-type star (B5-B9) from the OGLE colours. The light curve shows fluctuations of the order of 0.05 mag 
in the I band and evidence for a strong modulation at a period of 2.19\,days. The folded light curve appears sinusoidal which is probably evidence 
for non-radial pulsations in the star \citep{2008A&A...480..179D}.

The position, optical magnitudes and colours make the star closest to the X-ray position (OGLE-III 14642) the most likely counterpart.

\begin{table*}
\caption{Possible optical counterparts of \xmmp.}
\begin{center}
\begin{tabular}{cclcccccll}
\hline\hline\noalign{\smallskip}
\multicolumn{1}{c}{RA(2000)$^{(a)}$} &	
\multicolumn{1}{c}{DEC(2000)$^{(a)}$} &	
\multicolumn{1}{c}{dist.$^{(b)}$}&
\multicolumn{1}{c}{U (mag)$^{(a)}$} &
\multicolumn{1}{c}{B (mag)$^{(a)}$} &   
\multicolumn{1}{c}{V (mag)$^{(a)}$} &	
\multicolumn{1}{c}{I (mag)$^{(a)}$} &
\multicolumn{1}{c}{Q (mag)$^{(c)}$} &
\multicolumn{1}{c}{OGLE-III} &
\multicolumn{1}{c}{MACHO} \\
\noalign{\smallskip}\hline\noalign{\smallskip}
 00 48 14.10 & -73 22 03.6   &  0\farcs76  &  13.96$\pm$0.03 & 14.90$\pm$0.02 & 15.02$\pm$0.05 & 15.25$\pm$0.13 & -0.85$\pm$0.05 & 14642  & 212.15846.31   \\
 00 48 13.52 & -73 22 02.8   &  2\farcs86  &  $-$            & 18.90$\pm$0.07 & 17.26$\pm$0.07 & 15.88$\pm$0.04 & $-$            & 14688  & 212.15846.83   \\
 00 48 13.78 & -73 22 00.5   &  4\farcs07  &  15.20$\pm$0.03 & 15.93$\pm$0.03 & 15.90$\pm$0.03 & 16.00$\pm$0.04 & -0.75$\pm$0.05 & 14689  & 212.15846.70   \\
\noalign{\smallskip}\hline\noalign{\smallskip}
\multicolumn{9}{l}{\hbox to 0pt{\parbox{180mm}{\footnotesize
\smallskip
$^{(a)}$according to \citet{2002AJ....123..855Z}. \\
$^{(b)}$distance of the \citet{2002AJ....123..855Z} positions to the bore-sight corrected \xmm position. \\
$^{(c)}$Reddening-free Q-index = (U-B) $-$ 0.72 $\times$ (B-V).
}}}
\end{tabular}
\end{center}
\label{tab:opt-cp}
\end{table*}

\begin{table}
\caption{Optical and IR photometry of OGLE-III 14642}
\begin{tabular}{ccccc}
\hline\hline\noalign{\smallskip}
   & M2002$^{(a)}$  & K2007$^{(b)}$   & FTS$^{(c)}$   & Sirius$^{(c)}$  \\
   & 1998 Jan 8     & 2002 Aug 31     & 2009 Nov 25   & 2009 Dec 15 \\
   & MJD\,51186     & MJD\,52517      & MJD\,55160    & MJD\,55180  \\
\noalign{\smallskip}\hline\noalign{\smallskip}
  B & 14.54$\pm$0.01 & $-$ & 14.87$\pm$0.20 & $-$ \\
  V & 14.66$\pm$0.01 & $-$ & 14.86$\pm$0.03 & $-$ \\
  R & 14.66$\pm$0.01 & $-$ & 14.87$\pm$0.03 & $-$ \\
  I & $-$ & 14.70$\pm$0.01 & 15.30$\pm$0.02 & $-$ \\
  J & $-$ & 14.67$\pm$0.01 & $-$ & 15.46$\pm$0.01 \\
  H & $-$ & 14.62$\pm$0.01 & $-$ & 15.47$\pm$0.02 \\
  K & $-$ & 14.53$\pm$0.02 & $-$ & 15.49$\pm$0.07 \\
\noalign{\smallskip}\hline\noalign{\smallskip}
\multicolumn{5}{l}{\hbox to 0pt{\parbox{180mm}{\footnotesize
\smallskip
$^{(a)}$\citet{2002ApJS..141...81M}. 
$^{(b)}$\citet{2007PASJ...59..615K}. 
$^{(c)}$This work.
}}}
\end{tabular}
\label{tab:opt-mag}
\end{table}

\subsection{Long-term variability of OGLE-III 14642}

\begin{figure}
  \resizebox{\hsize}{!}{\includegraphics[width=0.1\textwidth,angle=0,clip=]{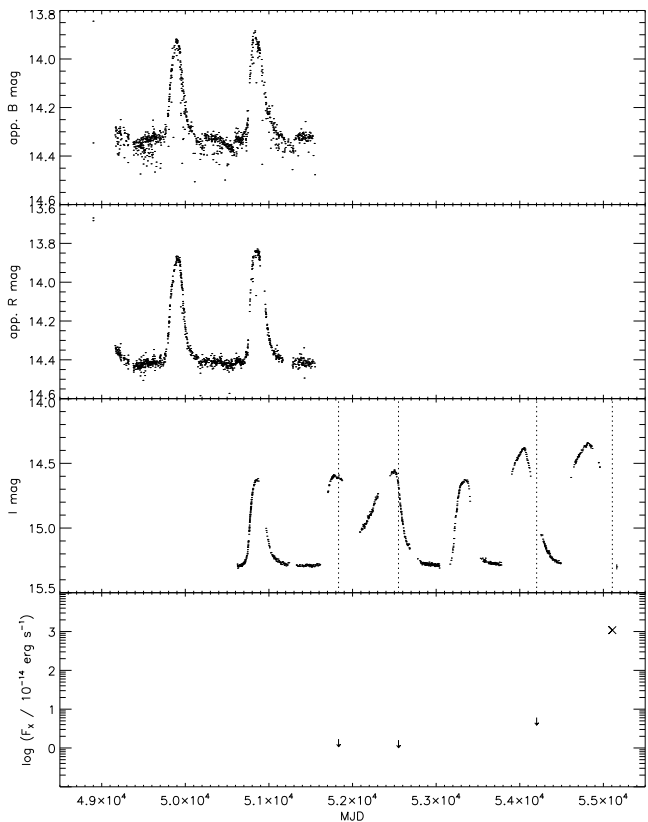}}
  \caption{Multi-wavelength light curves of the \xmmp\ / OGLE-III 14642 system. The upper two panels show the MACHO B- and R-band. 
           In the third panel the OGLE-III I-band light curve is plotted, with the last data point indicating our 
	   own measurement using the Faulkes telescope (see text). Dashed lines indicate the times of X-ray measurements, as 
	   shown in the bottom panel. Arrows mark upper limits (\xmm, Chandra and \xmm in chronological order, see Sect. 3.3), the cross 
	   indicates the XMM-Newton detection.}
  \label{fig:lc}
\end{figure}

The identification of \xmmp with OGLE-III 14642 is supported by the MACHO and OGLE light curves. This star shows strong 
outbursts repeating on time scales of $\sim$1000 days.
Fig.~\ref{fig:lc} shows the light curves of the proposed optical counterpart in approximate B and R magnitudes derived 
from MACHO data (ID 212.15846.31) and in the I-band from  OGLE-II and OGLE-III.
The I-band data point which we obtained at FTS was added to the OGLE light curve.
The MACHO light curve shows two outbursts around April 1995 and Jan. 1998, while OGLE observed six consecutive outbursts between June 1997 and 
Dec. 2008.

The X-ray flux of the source is not obviously correlated with the optical outburst activity.  The first \xmm non-detection was during maximum optical brightness.
The Chandra non-detection was close to the maximum optical brightness, the last \xmm non-detection later in the optical decline, while the detection could have 
happened at the decline or already in optical low-state (see Fig.~\ref{fig:lc}).

\subsection{Optical spectrum}

\begin{figure}
  \resizebox{\hsize}{!}{\includegraphics[clip=,angle=-90]{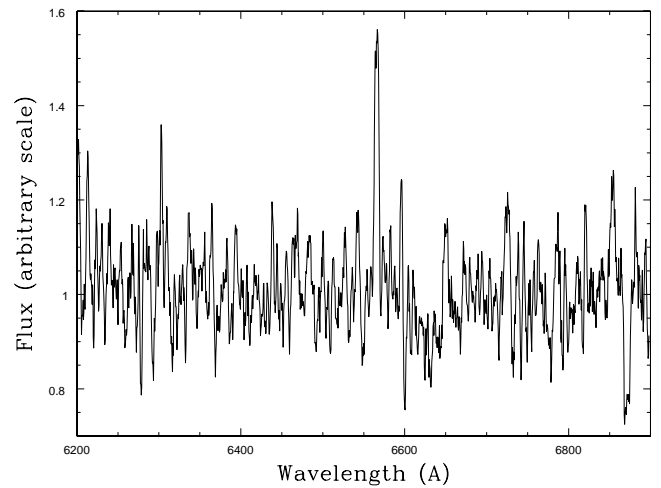}}
  \caption{H$\alpha$ spectrum of the OGLE optical candidate \#14642 taken 11 Dec. 2009 at SAAO.}
  \label{fig:halpha}
\end{figure}

Spectroscopic observations of the H$\alpha$ region were made on 11 Dec. 2009 (MJD 55176) using the 1.9\,m telescope of the 
South African Astronomical Observatory (SAAO). A 1200
lines per mm reflection grating blazed at 6800\,\AA\ was used with the SITe CCD which is effectively 266$\times$1798 pixels in
size, creating a wavelength coverage of 6200\,\AA\ to 6900\,\AA. The pixel scale in this mode was 0.42\,\AA/pixel.
The data were reduced using IRAF standard routines and the resulting
spectrum is shown in Fig.~\ref{fig:halpha}. 
The peak is at 6566\,\AA\ which is consistent with the corresponding rest wavelength of the \Halp\ line corrected for the motion of the SMC.
In this mode the spectral resolution is $\sim$0.2 nm for the signal to noise of $\sim$ 10.
We measured an H$\alpha$ line emission width of EW = 3.5 $\pm$ 0.6\,\AA, 
the error being calculated using the prescription given in \citet{1986MNRAS.222..809H}.

\section{Discussion and Conclusions}

One of the first \xmm observations of the SMC survey revealed the new high mass X-ray binary pulsar \xmmp with a pulse period of 11.866\,s 
\citep[following ][ we give it the alternative name SXP11.87]{2005MNRAS.356..502C}. 
Its X-ray behaviour and the properties of the optical counterpart (star with OGLE-III ID 14688) are typical for a Be - neutron star binary system.
In particular the appearance as an X-ray transient (a factor of at least 560 brighter during the outburst on Oct. 2009 as compared to non-detections 
from archival \xmm and Chandra observations), the hard power-law shape of the X-ray spectrum, the pulse period, the optical brightness, variability and colours
(indicating an early B star) and finally the H$\alpha$ emission line in the optical spectrum clearly confirm \xmmp as another Be/X-ray binary in the SMC.

The power-law photon index derived from the EPIC spectra of 0.53$-$0.66 (depending on the spectral modelling of the soft part of the spectrum by including 
an additional soft model component or not) is on the hard side of the distribution of photon indices for Be/X-ray binaries in the SMC, 
which shows a maximum at $\sim$0.9$-$1.0 \citep{2008A&A...489..327H,2004A&A...414..667H}. The range of values obtained for \xmmp 
is similar to the index of 0.35$-$0.54 reported for the 6.85\,s pulsar XTE\,J0103$-$728 
\citep[also using \xmm data in the same energy band;][]{2008A&A...484..451H}. It should be noted, 
that the latter, also called SXP6.85, was detected at energies up to 35 keV with RXTE and INTEGRAL during a long type II outburst \citep{2010MNRAS.403.1239T} 
showing that the hard spectrum extends to energies beyond the sensitivity of \xmm. SXP11.87 and SXP6.85 also show similarities in 
their spectra at energies below 2 keV, indicating a soft X-ray excess. However, the energy resolution of the CCD instruments is not 
sufficient to determine the exact nature of this component. Some constraints can be inferred by using different models for the soft component.
If one assumes a black-body component, 
a temperature of $\sim$280\,eV and black-body radius of $\sim$13.4 km is derived for SXP11.87. While this could still be compatible with the size of the 
neutron star, the corresponding black-body radius for SXP6.85 is too large \citep[30 km,][]{2008A&A...484..451H}. 
Therefore, these authors conclude that the soft excess more likely originates near the inner edge of an accretion disk as expected 
for intermediate X-ray luminosities. The very similar 
parameters derived for a soft excess emission suggest the same picture for SXP11.87, although the inferred black-body or inner disk radii 
seem to be smaller than the corresponding values for SXP6.85. However, also
emission by diffuse gas through collisional heating or photoionisation is possible for both cases \citep{2004ApJ...614..881H}.

The MACHO and OGLE light curves of the optical counterpart of \xmmp show prominent outbursts repeating on a time scale of about 1000 days. 
Very similar behaviour was reported from the optical counterpart of the 18.37\,s Be/X-ray binary pulsar XMMU\,J004911.4-724939 which showed
two outbursts separated by about 1300 days in MACHO and OGLE-I data \citep{2008A&A...489..327H}. Such outburst behaviour is also observed 
from other (single) Be stars \citep{2002A&A...393..887M}. Because of this and the fact that the outbursts do not repeat strictly periodically, 
it is unlikely that they are related to the orbital period of the binary system. 
Moreover, from the Corbet relation between neutron star spin period and the orbital period \citep{1984A&A...141...91C} a much shorter orbital period of 
about 20$-$200 days is expected \citep[see][ for more recent versions of the P$_{\rm s}$/P$_{\rm orb}$ diagram]{2005ApJS..161...96L,2009IAUS..256..361C}.

\begin{acknowledgements}
The XMM-Newton project is supported by the Bundesministerium f\"ur Wirtschaft und 
Technologie/Deutsches Zentrum f\"ur Luft- und Raumfahrt (BMWI/DLR, FKZ 50 OX 0001)
and the Max-Planck Society. 
R.S. acknowledges support from the BMWI/DLR grant FKZ 50 OR 0907.
S.M., N.L. and A.T. acknowledge the support of ASI through contract I/088/06/0.
L.J.T. is in receipt of a University of Southampton Mayflower Scholarship.
A.U. acknowledges support from the MNiSW/BST grant.
\end{acknowledgements}

\bibliographystyle{aa}
\bibliography{../general}

\end{document}